# Carrier Transport at the Metal-MoS$_2$ Interface

Faisal Ahmed[1,2], Min Sup Choi[1,3], Xiaochi Liu[1,3] and Won Jong Yoo [1,2.3, *]

[1]Samsung-SKKU Graphene Center (SSGC), SKKU Advanced Institute of Nano-Technology (SAINT), Sungkyunkwan University, 2066, Seobu-ro, Jangan-gu, Suwon, Gyeonggi-do, 440-746, Korea

[2]School of Mechanical Engineering, Sungkyunkwan University, 2066, Seobu-ro, Jangan-gu, Suwon, Gyeonggi-do, 440-746, Korea

[3]Department of Nano Science and Technology, SKKU Advanced Institute of Nano-Technology (SAINT), Sungkyunkwan University, 2066, Seobu-ro, Jangan-gu, Suwon, Gyeonggi-do, 440-746, Korea

*Corresponding email: yoowj@skku.edu

*Abstract*

This study illustrates the nature of electronic transport and its transition from one mechanism to another between a metal electrode and MoS$_2$ channel interface in a field effect transistor (FET) device. Interestingly, measurements of the contact resistance ($R_c$) as a function of temperature indicate a transition in the carrier transport across the energy barrier from a thermionic emission at a high temperature to tunneling at a low temperature. Furthermore, at a low temperature, the nature of the tunneling behavior is ascertained by the current-voltage dependency that helps us feature direct tunneling at a low bias and Fowler-Nordheim tunneling at a high bias for a Pd-MoS$_2$ contact due to the effective barrier shape modulation by biasing. In contrast, only direct tunneling is observed for a Cr-MoS$_2$ contact over the entire applied bias range. In addition, simple analytical calculations were carried out to extract $R_c$ at the gating range, and the results are consistent with the experimental data. Our results describe the transition in carrier transport mechanisms across a metal-MoS$_2$ interface, and this information provides guidance for the design of future flexible, transparent electronic devices based on 2-dimensional materials.

**Introduction**

In recent years, transition metal dichalcogenides (TMDCs) have garnered a tremendous amount of attention from the research community due to their splendid properties, and two-dimensional layered $MoS_2$ is a leading material in the TMDC family as a result of its ultra-thin body, absence of dangling bonds and large band gap.[1,2] These unusual properties make it a promising material with potential uses in electronics,[3,4] optical devices[5,6] and memories[7]. It is also a promising channel material for use in field effect transistor devices because it has a mobility in the hundreds, superb on/off ratio of about $10^7 \sim 10^8$ and low subthreshold swing of around 74 mV/decade.[1,2,8] The resilience to the short channel effect, quantum confinement in the channel, mechanical flexibility and suppressed surface scattering due to its ultra flat surface show that $MoS_2$ based devices have superior properties when compared to conventional Si technology. The pristine surface of $MoS_2$ offers no dangling bonds, enabling a weak Van der Waals contact to be induced when a metal is deposited over the top of it. Therefore, unlike conventional devices, the basic operation of a two-dimensional $MoS_2$ device is dominated by the properties of the contacts.[9-11] Thus, it is essential to have a solid understanding of contact engineering to fabricate efficient $MoS_2$ devices. In general, when $MoS_2$ comes into contact with certain metals, a Schottky barrier forms at the interface due to the mismatch in the work function, giving rise to contact resistance ($R_c$). The magnitude of $R_c$ depends on the nature of the barrier, *i.e.* its width and height, since the barrier aggressively affects carrier transport across it. Few reports have attempted to optimize metal-$MoS_2$ contacts to ensure efficient charge injection, and two main approaches have been adapted to this end: reducing the Schottky barrier height[12-15] or thinning the barrier width.[16-19] Liu *et al.* studied the change in a metal-$MoS_2$ Schottky barrier (SB) with respect to biasing as well as its impact on rectification,[10] and in our previous report, we illustrated the bias effect on SB modulation to harness an efficient photo response.[20] Das *et al.* calculated Schottky barrier heights between different metal electrodes and the $MoS_2$ channel,[15] and in another report, they also explained carrier distribution and transport across different layers of the $MoS_2$ channel.[21] However, an in-depth study on the nature of charge carrier transport along the interface between a metal electrode and a $MoS_2$ channel is still lacking. Some basic questions are yet to be answered. What type of carriers is dominant at certain conditions along the interface? When does the transition from one mechanism to the other occur? How much do they contribute to $R_c$?

In this study, we have tried to bridge the gap by systematically elaborating on the different carrier transport mechanisms that are involved along the interface. We carried out low temperature measurements on contact properties of the $MoS_2$ devices. As a result, different behaviors of charge injection across the interfacial barrier and their cross over were clearly visualized. In order to further elaborate the analysis, we measured the $R_c$ of the metal-$MoS_2$ junction as a function of the temperature to examine the competition between thermionic emission and tunneling transport at the interface. In addition, we also investigated the nature of the tunneling behavior by using the simplified mathematical models for Fowler Nordheim (F-N) tunneling and direct tunneling. We found that for Pd-$MoS_2$, an obvious transition is observed from direct to F-N tunneling. In contrast, only direct tunneling occurs for Cr-$MoS_2$. Finally, we used the Landauer theory[22,23] to analytically calculate $R_c$ contributed by the current components and combined them to obtain the net $R_c$ value, which we found to be consistent with the experimental results.

**Experimental Details**

For our experiment, few layers MoS$_2$ flakes were mechanically exfoliated using scotch tape onto a *p*-type Si substrate capped with thermally grown 285 nm SiO$_2$ that served as the global back gate. The substrate was baked on a hot plate at 100°C for 10 min before exfoliation in order to remove water molecules from the surface. The electrodes were patterned via electron beam lithography (EBL) following the transmission line method (TLM) [See Figs. 1(a), 1(b)] to extract the gate-modulated $R_c$. Two different metallizations with 5/50 nm of Cr/Au and 10/40 nm of Pd/Au were carried out via electron beam deposition. Cr and Pd were selected since they form lower and higher SBs with respect to MoS$_2$,[6] as we can study the dependence of barriers on carrier transport. Note that only the results for the Cr/Au-deposited devices are shown unless otherwise mentioned.

A semiconductor parameter analyzer was used to carry out the electrical measurements, and the low-temperature measurements were performed from room temperature down to 120K by using liquid nitrogen. $R_c$ was extracted at a given number of temperature points and the range of the gating to further detail the behavior of $R_c$ as a function of temperature as shown in Fig. 1(c). Further details about calculation of $R_c$ can be found in our previous study.[24]

**Results and Discussion**

*$R_c$ vs. T*

Figure 1(c) reveals an increase in $R_c$ as the temperature falls from room temperature to 123K. These results can be further described by dividing the graph into two temperature regions with a high temperature region from 298K to 248K and a low temperature region from 248K to onwards. In the high temperature region, the increase in $R_c$ as the temperature decreases is quicker than in the low-temperature region. This behavior can be explained by considering the carrier transport at the interface. At the metal(Cr)-MoS$_2$ Schottky contact, a charge injection occurs either (i) as a result of thermionic emission over the top of the barrier due to the transfer of thermal energy from phonons to electrons to surmount the barrier height or (ii) as a result of quantum mechanical tunneling of carriers across the barrier width. In fact these transport mechanisms have different sensitivities to temperature,[25,26] and thus their respective $R_c$ vary as temperature varies. In the high temperature region, thermionic emission, which is readily temperature sensitive, is the dominant transport mechanism across the interface [See Fig. 1(d)]. Thus, a slight fall in the temperature drastically suppresses the thermionic current and sufficiently increases $R_c$. However, in the low temperature region, tunneling seems to be the dominant transport mechanism across the interface [See Fig. 1(d)], since tunneling is less sensitive to temperature and the change in $R_c$ is very small. This small change in $R_c$ can be attributed to the suppression of thermally-assisted tunneling across the barrier due to further cooling. Similar results were also obtained for the Pd-MoS$_2$ contact, as shown in Fig. S1. Fig. 1(c) also indicates that the same trend for $R_c$ with respect to temperature is observed for all the gate voltages that we measured, but the increase in $R_c$ in the first region is less pronounced as the gate voltage increases, indicating that thermionic emission is suppressed by additional gating. When a higher gate bias is applied, the energy levels of MoS$_2$ are pulled down that leads to a thinning of the interfacial barrier and an increase in the tunneling probability of the carriers, resulting in enhanced tunneling current or in other words the channel is electro-statically doped. Thus we observe very little modulation in $R_c$ with respect to temperature at a high bias since carrier transport is dominated by tunneling. This means increasing gate bias shifts transition point towards high temperature. The plot of $R_c$ as a function of the temperature is conclusively the hallmark that clearly differentiates the dominant transport mechanisms at certain points across the barrier.

Unlike for metal-MoS$_2$, the $R_c$ at the metal (Pd)-graphene interface declines as the device cools.[27] This contradictory temperature dependency is mainly a result of a difference in the origin of $R_c$ along these two junctions. Graphene under a metal electrode is more responsible for $R_c$ in the metal-graphene interface. When the temperature decreases, the carrier transport across the interface changes from diffusive to ballistic, mainly due to coupling length and the carrier mean free path, that eventually suppresses $R_c$. This explains why the $R_c$ of pure edge-contacted graphene shows no variation with temperature.[28] However, a metal-MoS$_2$ contact, as explained in the previous paragraph, has an $R_c$ that originates from the formation of the barrier, and its temperature sensitivity depends on the carrier transport across it.

*Tunneling Behavior*

As mentioned earlier, tunneling is the dominant mechanism for charge transport across the barrier at low temperature. The tunneling behavior can be direct or Fowler-Nordheim (F-N) depending on shape and width of barrier. But which occurs at a given point? To answer this question, we use direct and the F-N tunneling equations (1a and 2a) and mathematically test the linearity of the data using the equations (1b and 2b) for easy comparison.[29,30]

Direct Tunneling

$$I \propto V \exp[-\frac{4\pi d \sqrt{2m^* \phi_B}}{h}] \quad (1a)$$

$$\ln(\frac{I}{V^2}) \propto \ln(\frac{1}{V}) - \frac{4\pi d \sqrt{2m^* \phi_B}}{h} \quad (1b)$$

Fowler−Nordheim Tunneling

$$I \propto V^2 \exp[-\frac{8\pi d \sqrt{2m^* \phi_B^3}}{3hqV}] \quad (2a)$$

$$\ln(\frac{I}{V^2}) \propto -\frac{1}{V}(\frac{8\pi d \sqrt{2m^* \phi_B^3}}{3hq}) \quad (2b)$$

Here $\phi_B$ is the barrier height, $m$ is free electron mass, $m^*$ $(0.46m)^{31}$ is the effective mass of electrons in the MoS$_2$ channel, $q$ is the electron charge, $h$ is Planck's constant and $d$ is the width of the barrier.

Equations (1b) and (2b) imply that direct and F-N tunnelings differ in terms of I-V dependency. Therefore if the plot for $ln(I/V^2)$ vs. $1/V$ shows linearity, then F-N tunneling is expected to occur, whereas when the slope rises exponentially, direct tunneling is thought to occur. The main graph in Fig. 2(a) displays an almost exponential plot throughout the applied bias range, which indicates direct tunneling is the dominant mechanism for the Cr-MoS$_2$ contact. The inset in the same graph, which is plotted according to Equation (1b), shows a linear trend that further confirms the direct tunneling. In contrast, Fig. 2(b) shows that, for the Pd-MoS$_2$ contact in the high bias region (left side of the graph), a linear decrease first reaches a specific point and then rises exponentially in the low bias region, which reveals a transition from F-N (colored area) to direct tunneling. In order to explain this anomaly, we investigate the band diagram along the interface of both contacts. The direct tunneling and the F-N tunneling are determined by the nature of interfacial barrier, that is, the former occurs when the barrier is trapezoidal (wide) and the latter occurs when the barrier is triangular (thin).[29,30] Generally, a MoS$_2$ device has two contacts that induce their respective SBs: the source SB and the drain SB. The shape, width and height of these barriers are mainly modulated by applied bias,[10,20] affecting the carrier injection behavior. First, consider the Pd-MoS$_2$ contact [Fig. 2b]. When a high drain bias is applied, the drain barrier reduces and eventually vanishes but the source barrier becomes thin. Therefore, at a low drain bias the carriers have to overcome two wide barriers so the direct tunneling is realized, whereas at a high drain bias they only experience a thin and triangular source barrier that favors F-N tunneling. As the result, the change in the transport mechanism from direct tunneling at the low drain bias to F-N tunneling at the higher drain bias is realized at the Pd-MoS$_2$ interface [Fig. 2(d)]. This crossover occurs at around 0.22V (4.5V$^{-1}$), and it is worth noting here that as the temperature increases from 123K to higher temperatures, the amount of F-N tunneling that occurs keeps decreasing and completely vanishes at around room temperature. This observation is consistent with our earlier discussion in that the tunneling current is dominant mainly in the low temperature regime. In addition, we also extract the width of Pd-MoS$_2$ interface from F-N tunneling equation. By substituting the slope of linear portion of Fig. 2(b), SB height and effective mass of 0.25eV and 0.46m respectively,[31] in Equation (2b), the effective barrier width (d) of around 0.3nm is obtained for Pd-MoS$_2$ junction.

However for the Cr-MoS$_2$ contact, there is no sign of F-N tunneling throughout the applied bias sweep. One major difference between these two metals can be seen in their work functions. With respect to MoS$_2$ (4.2~4.6eV), Cr (4.6eV) has a lower work function whereas Pd (5.0eV) has a higher work function, so they form a lower and a higher SB height with MoS$_2$, respectively.[6] Besides barrier height, tunneling depends more severely on its width since the charged carriers have to tunnel quantum mechanically throughout the barrier width. Therefore, this anomaly could not be explained simply by considering the differences in the work function and the SB height. As mentioned earlier, MoS$_2$ contains pristine surface without dangling bonds. Therefore, when a metal is deposited over the surface of MoS$_2$, a weak Van der Waals interaction occurs between them, inducing a physical separation [tunnel barrier (TB)] along with the SB at the contacts. For example, the extent of TB depends partly on the difference of lattice structures between deposited metal and MoS$_2$. It is reported that Cr and MoS$_2$ have large mismatch in their lattice structure, whereas this difference is very small between Pd and MoS$_2$.[11] Therefore, when MoS$_2$ comes into contact with Cr, a weak overlapping occurs in their orbitals that induce a wide TB at their interface along with SB as shown in Fig. 2(c). On the other end, the better orbital overlapping and a narrow TB is observed at the Pd-MoS$_2$ junction [Fig. 2 (d)]. Besides physical mismatch, the unique properties of metals with respect to MoS$_2$ may also partly affect the nature of TB. We think that due to high chemical reactivity of Cr, the partial oxidation of Cr might occur due to uninvited surface contaminations introduced during EBL process that may further induce wide TB at Cr-MoS$_2$ interface. Moreover, Pd has better wetting ability to MoS$_2$ surface and a uniform growth of Pd is also expected, that may also cause a narrow TB at their junction.[32] As explained in previous paragraph, by applying a high drain bias, the drain SB is vanished and the source SB is thinned, but the TBs may remain intact from these changes due to its physical nature. Therefore, at a high voltage the effective barrier width still remains wide for the Cr contact, but it is thinned for the Pd contact since it is mainly dominated by TB for the former and by the SB for the latter contact. As a result, we observe only direct tunneling without realizing F-N tunneling at the Cr-MoS$_2$ contact, but a clear transition is observed from one behavior to another at Pd-MoS$_2$ interface.

*Analytical calculation of $R_c$*

In addition to the experimental measurements, numerical calculations were carried out to extract $R_c$ across the metal (Cr)-MoS$_2$ interface theoretically. A simple scheme is proposed to extract the $R_c$. We used the well defined analytical carrier transport model proposed by Das *et al.* for metal-MoS$_2$ interface,[31] and which is also successfully advanced to metal-phospherene junction recently.[33] By implementing that model, current components shown in the band diagram of Fig. 3(a) across the interface are calculated. We applied the classical Landauer theory; $R_c=h/(2q^2MT)$, where $h$ is Planck's constant, $q$ is the electron charge, $M$ is the number of conduction modes in MoS$_2$ channel and $T$ is the transmission probability of carriers,[22,23,27] to the extracted current components in order to estimate their respective $R_c$. Finally, these all the $R_c$ components are combined by a simple electrical model to extract their total $R_c$. Interestingly, the $R_c$ estimated by adopting our scheme is consistent with the experimental results across the range of applied gate bias. The readers should note that ballistic transport in the channel is assumed in Landauer theory so channel resistance is underestimated in our calculations. However, this assumption could be justified from the fact that MoS$_2$ device operation is much dominated by contacts rather than channel of the device. The similar assumption was also made in previous reports.[27,31,33]

Generally, carriers along the metal-MoS$_2$ interface are divided into three components *i.e.* thermionic emission ($I_{TH}$) over the top of barrier and tunneling components ($I_{TN-1}$ and $I_{TN-2}$) along their respective regions as depicted in the energy band diagram of Fig. 3(a). The numerical equations of all three current components along with their detailed calculations procedures are illustrated in Supporting Information S2 and their results are shown in Fig. 3(b) in the units of A/m. All the current components are gate dependent and can be explained by the barrier modulation theory. The thermionic emission ($I_{TH}$) current component increases due to the decrease of the effective barrier height, and the tunneling components ($I_{TN-1}$ and $I_{TN-2}$) enhance because of thinning of the effective barrier width, respectively, when more gate bias is applied. Next, $R_c$ of each current component is extracted by applying the simplified Landauer formula *i.e.* $R_c=1/I \cdot q$, where $I$ is the current component and $q$ is the electron charge,[22,23] to the current components, since the applied drain bias is one volt therefore the chemical potential difference becomes unity. Their result is shown in Fig. 3(c) after normalizing to the standard units of $R_c$ *i.e.* (ohm·mm). As expected, the current component with smaller magnitude across the barrier contributes significantly to the $R_c$ at the same bias condition. As mentioned earlier the carriers across the interface split into three parallel paths (See band diagram) so we replace Fig. 3(a) with a parallel electrical resistor network shown in Fig. 3(d) to combine all $R_c$ values. Finally, their net result is shown in Fig. 3(e) and compared with experimentally calculated results of $R_c$. Note that $I_{TN-2}$ current level is very low i.e. around $10^{-32}$ A/m at 70 gate bias and its corresponding $R_c$ is extremely large i.e. around $10^{31}$ ohm·m, not shown in Fig. 3(c), that is much higher than acceptable range of $R_c$. Interestingly, after applying the proposed model the extracted total $R_c$ value is within the acceptable range and agrees well with our experimental results. However at low gate bias where device is near off-state, the difference between theoretical and experimental data is little bit large and the gap is reduced as the device enters into strong accumulation region.

However, the difference between the two results could be attributed to the assumption made during analytical calculation. Interestingly, despite this, the analytically calculated $R_c$ values in our scheme sweep to several ohm·mm depending on gate bias which are close to the experimentally measured $R_c$ for metal-MoS$_2$ interface by other groups.[14,17-19] Conclusively, using the proposed model above one can easily calculate $R_c$ across the range of gate bias for metal-MoS$_2$ interfaces.

Currently, the lowest reported value for $R_c$ in a metal-MoS$_2$ contact is still several orders of magnitude higher than the acceptable levels for miniaturized electronics.[34] However, by adopting the carrier transport techniques illustrated in this report, one can effectively reduce the $R_c$ values to appreciable limits, such as by (i) selecting an appropriate metal, which will preferably have a lower work function and an effective orbital overlapping with MoS$_2$, since this will discourage SB and TB and will enhance thermionic emission and tunneling across the barrier; (ii) doping the contact region since a degenerate and stable doping technique can induce a much thinner barrier that will facilitate carriers to tunnel through it; and (iii) using an edge contact since it has been theoretically proposed that an edge contact more efficiently injects the carriers than a surface contact for TMDCs due to their layered body.[11] Carefully controlling the edge etching and the defects can produce a one-dimensional contact for $MoS_2$. Above all, the techniques solely depend on carrier injection, thus fundamental knowledge on carrier injection will be helpful to achieve optimum contacts.

In summary, the temperature-dependent carrier transport in a metal-$MoS_2$ interface was systematically investigated according to several charge injection mechanisms and their transitions. The transition from thermionic emission to tunneling was observed at around 248K temperature. In addition, an anomaly in terms of differences in the tunneling behavior was spotted for Cr-$MoS_2$ and Pd-$MoS_2$ contacts, which suggests a difference in the nature of the barrier that formed along the interface. This work is a promising approach towards realizing optimized metal-$MoS_2$ contacts for future devices using 2-dimensional materials.

**Acknowledgements**

This research was supported by the Basic Science Research Program through the National Research Foundation of Korea (NRF) (2013-015516) and by Global Frontier Program through the Global Frontier Hybrid Interface Materials (GFHIM) of NRF funded by the Ministry of Science, ICT & Future Planning (2013M3A6B1078873).

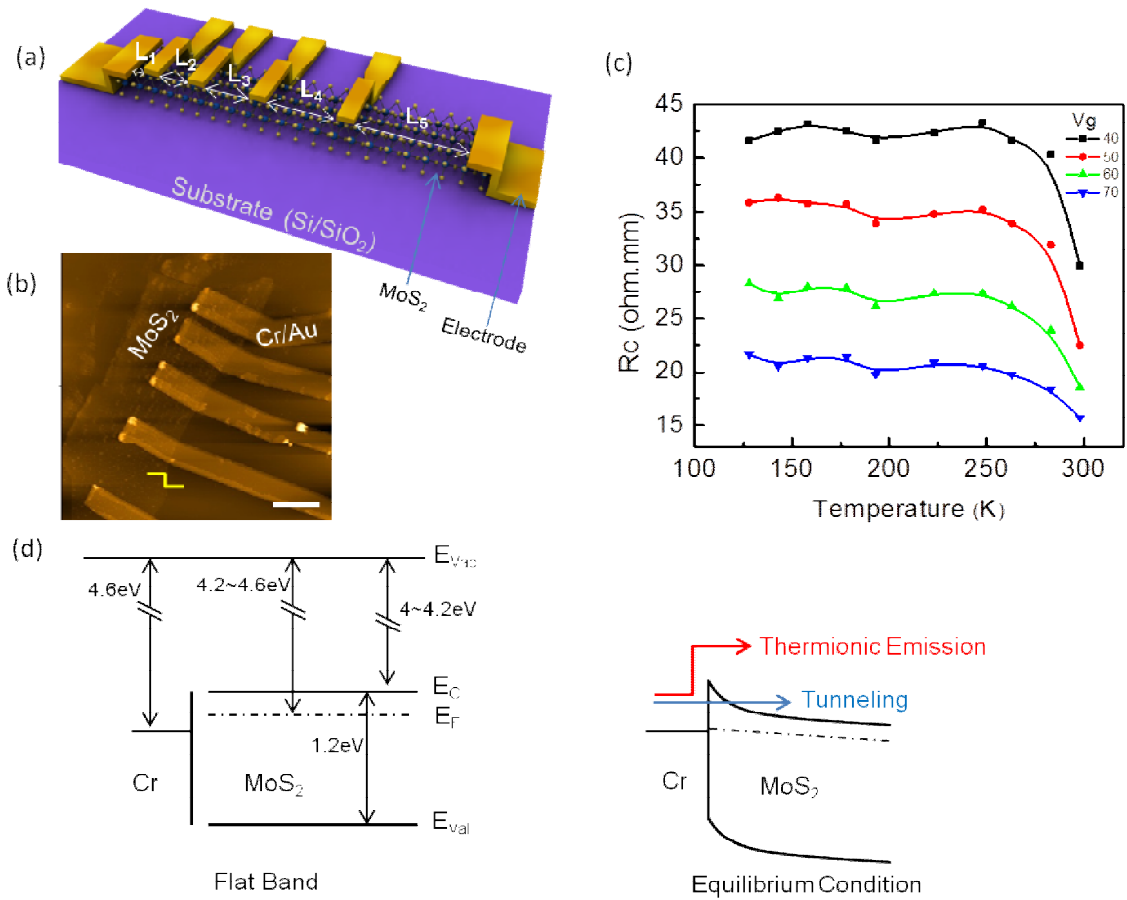

**Figure 1 (a)** Schematic of the TLM-patterned $MoS_2$ device where $L_5>L_4>L_3>L_2>L_1$ **(b)** AFM image of the TLM device with scale bar of 2 μm, where yellow step denotes the flake thickness of around 14 nm and the channel lengths from $L_1$ to $L_4$ are 0.92, 1.45, 1.97 and 2.47 μm respectively. The channel width is of 3.5μm **(c)** $R_c$ vs. $T$ plot for the Cr-$MoS_2$ device at a given gate bias and drain bias is swept from -1 to 1V during output curve measurement. The points represent measured values and lines are guide to eyes. **(d)** Band diagrams of the device where the left side represents the flat band condition and the right side represents the equilibrium condition after the contact is made.

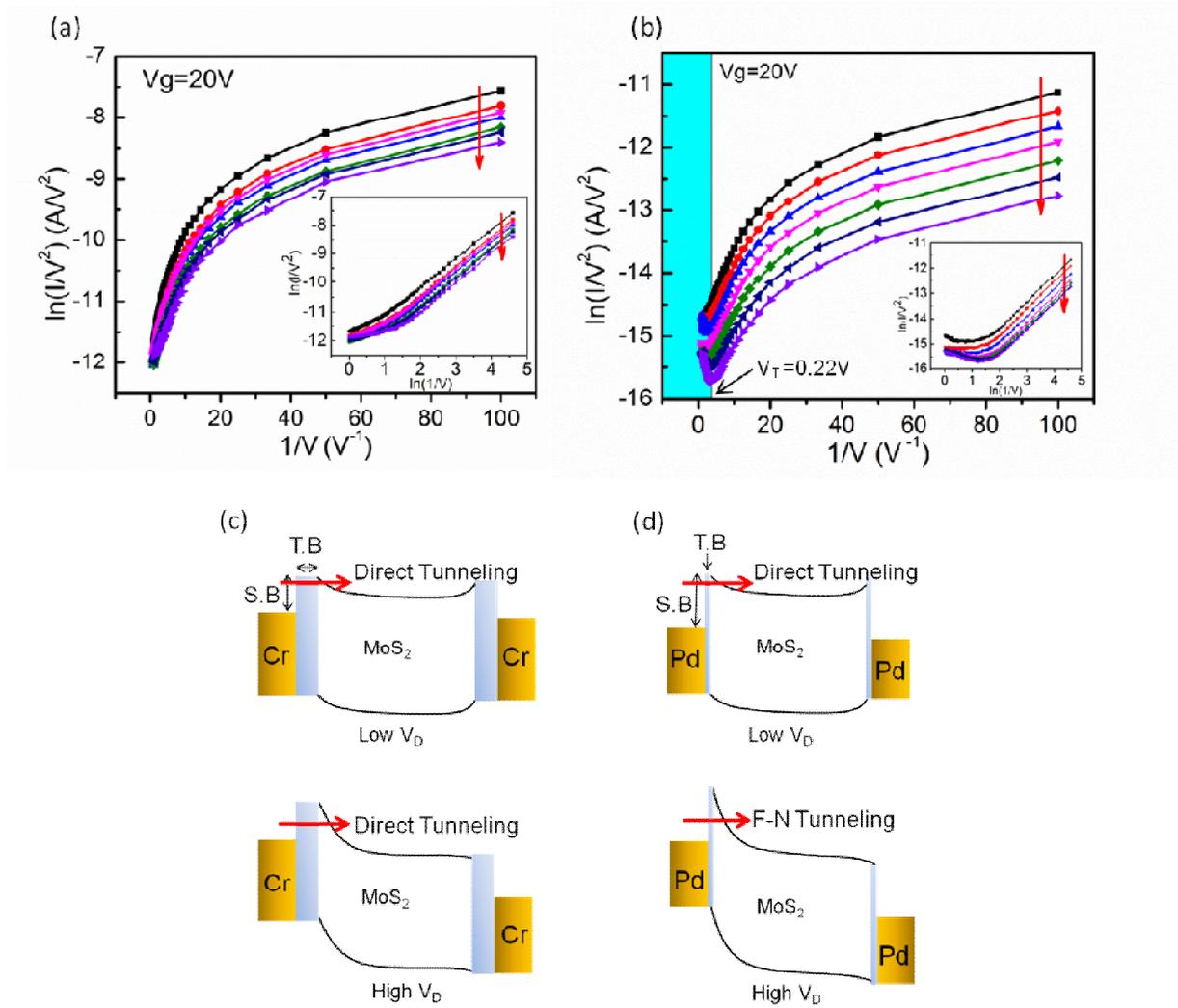

**Figure 2** $\ln(I/V^2)$ plotted vs. the inverse of the drain bias (1/V) **(a)** for the Cr contact and **(b)** for the Pd contact. Inset shows the same plot with logarithmic abscissa. The arrow denotes a decrease in temperature from room temperature down to 123K. **(c)** and **(d)** are the band diagrams of (a), (b), where SB and TB denote a Schottky barrier and a tunnel barrier, respectively. Note that $0 \leq V_D \leq 1$

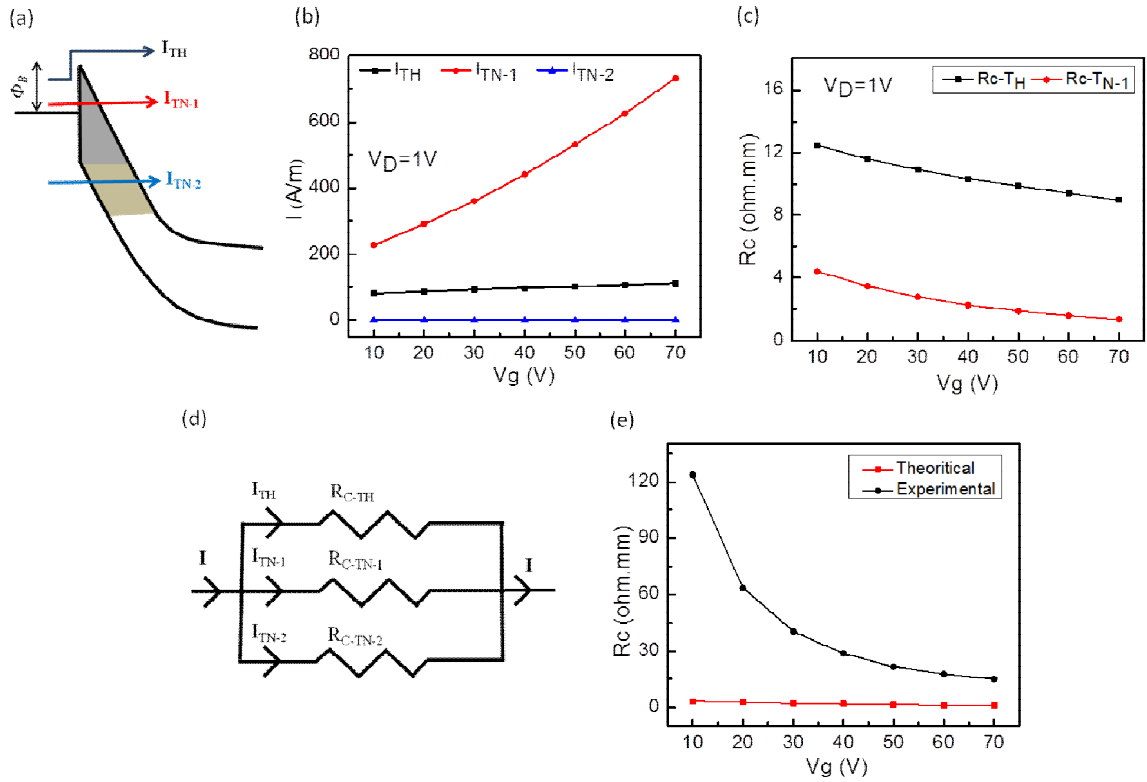

**Figure 3(a)** Band diagram of a metal-$MoS_2$ interface showing all three components, where $I_{TH}$ is the thermionic emission current, and $I_{TN-1}$ and $I_{TN-2}$ are the tunneling currents of their respective regions **(b)** Theoretically calculated current components, as shown in (a) in units of A/m. **(c)** The analytically calculated $R_c$ (in ohm·mm) for each of the current components. Note that $R_c$ of $I_{TN-2}$ component is not shown here since its value is too large. **(d)** The assumed parallel resistor network that replaces the band diagram of (a). **(e)** The combined result of all three resistance components measured in (c) according to (d) and compared against the experimental results.